\documentclass[twocolumn,pre,showpacs]{revtex4-1}
\usepackage{graphicx}

\begin{document}

\def\br{{\bf r}}
\def\vcmi{v_{cm,1,I}}
\def\pcmi{p_{cm,I}}
\def\vcm{{\bf v}_{cm}}

\title{Collision of Polymers in a Vacuum}

\author{J. M. Deutsch}
\affiliation{
Department of Physics, University of California, Santa Cruz, California 95064}

\date{\today}

\begin{abstract}
In a number of experimental situations, single polymer molecules can
be suspended in a vacuum.  Here collisions between such molecules
are considered. The limit of high collision velocity is investigated
numerically for a variety of conditions. The distribution of contact times,
scattering angles, and final velocities
are analyzed. In this limit, self avoiding chains
are found to become highly stretched as they collide with each other, and
have a distribution of scattering times that depends on the scattering
angle. The velocity of the molecules after the collisions is similar to
predictions of a model assuming thermal equilibration of molecules during
the collision. The most important difference is a significant subset of molecules that
inelastically scatter but do not substantially change direction.
\end{abstract}

\pacs{
}

\maketitle

\section{Introduction}

Although polymer molecules are most commonly studied in solution or in solid form~\cite{degennes}, there has been
increasing technological use for them in a vacuum. They are often prepared in this state as part
of the technique used to identify protein molecules with mass spectrometry~\cite{Hillenkamp}. 

Recently the properties of single chain molecules in a
vacuum were studied theoretically and by means of computer
simulation~\cite{DeutschVacPRL,DeutschExactVac,mossa,Taylor,DeutschCerf}.
It was shown that such molecules have unusual statistical properties
and that the dynamics are very different from those found for similar
molecules in solution. With no excluded volume, the lack of a solvent
means that the only damping that can occur is internal to the chain and
this leads to slowly damped oscillatory behavior for time dependent
correlation functions of polymer position. With excluded volume,
the time constant for relaxation scales with chain length $N$
as $N^{1.15 \pm .05}$ substantially faster than for corresponding
chains in solution~\cite{degennes}.  Oscillatory behavior is quite
pronounced for short chains~\cite{Taylor} but is suppressed when they
are longer~\cite{DeutschVacPRL}.

The equilibrium size of such chains is also influenced by the conservation
of angular momentum~\cite{DeutschExactVac}, and the exact statistical
properties of an ideal chain with this constraint can be calculated. When
the total angular momentum is zero, the radius of gyration is smaller
relative to a chain without this constraint enforced.

In experimental situations, these polymer chains are often charged and are
accelerated by an external field.  As far as the author is aware, there
have been no experiments that have measured their conformations in this
state. Such experiments would be very interesting and might allow for
the probing of additional features that could be used to characterize the
chemical composition of such chains. An important aspect in understanding
this situation are the nature of collisions between chains.

In equilibrium, collisions will obey detailed balance and the
average properties of two chains before and after a collision will be
identical. Under the conditions necessary to perform mass spectrometry,
nonequilibrium considerations become necessary.  Because the molecules
are charged and accelerated by large fields, they acquire a high center
of mass velocity. It seems likely that under such circumstances, one would
have collisions occurring between molecules with high relative
velocities, as their charges and masses are not all identical. Such
collisions differ from most collisions in that they are occurring between
very large molecules, and are expected to have quite a different character
than previously studied. 

In addition it should be possible to directly
study collisions between molecules with high relative center of mass velocities
using modifications to the highly sophisticated apparatus that is
in current use such as matrix-assisted laser desorption/ionization mass
spectrometry (MALDI MS). The acceleration voltage typically of order
order $10 kV$~\cite{BarnesChiu}. So with a single charge on a protein this amounts to a
center of mass kinetic energy of $10^4 eV$. Compared to the center
of mass thermal energy at $400 K$, this over $10^5$ times greater. 
It should be noted that this large field does not cause further ionization
of the molecules because the acceleration occurs over centimeters and so the
electric field is still small compared with that needed for ionization.
There are many sophisticated variations~\cite{MoonYoonKim} of MALDI that suggest
that modification of the apparatus for the purpose of studying collisions
should be feasible.

Therefore it is of interest to explore situations where the collision
velocities of molecules are much larger than the thermal values given
by the equipartition theorem~\cite{Reif} and this is done in this paper.

Throughout this work we will consider collisions in the center of mass frame. Transforming to
other frames is straightforward. For highly inelastic collisions, such as the ones described
below, this is the most natural reference frame to use. 

\section{The high collision velocity limit}

Self avoiding chains are characterized by a mass per monomer $m$,  step length $l$, and an excluded volume parameter.
Excluded volume can be understood as an effective hard core radius which prohibits monomers
from getting closer than a certain distance.  For the sake of generality, consider that the
chains have different numbers of monomers $N_1$ and $N_2$ and corresponding masses $M_1 = m N_1$ and $M_2 = m N_2$.
We will consider the collision in the center of mass reference frame, so that initially, the total momenta of the molecules
are ${\bf P}_I$ and $- {\bf P}_I$.
In the limit where $P_I$ is very large and the internal
energy of a chain is kept constant, the internal energy of a chain
becomes negligible in comparison with the center of mass kinetic energy. During a collision,
we will see that typically, a substantial fraction of this momentum is transfered into internal kinetic
energy of the chains. Therefore in this limit, the internal energy of a chain before a collision
can be neglected, and we can set the initial kinetic energy of the chains equal to zero.
In this case, with only hard core potentials, the scattering of self avoiding chains only depends
on velocity as a prefactor. The angular dependence of scattering becomes independent of $P_I$, and the
distribution of  the final scattered momentum ${\bf P}_F$, only depends on ${\bf P}_F/P_I$. This is an interesting limit to
consider because of the lack of dependence on the chains' temperature, and therefore this
will be studied in detail below.

\begin{figure}[htp]
\begin{center}
\includegraphics[width=\hsize]{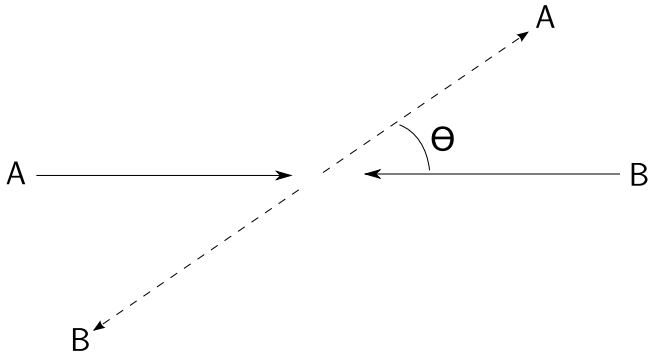}
\caption{ (Color Online) 
A collision in the center of mass frame. Polymers A and B move toward each other as shown
by the solid line. After the collision they come apart as shown by the dashed lines. The angle
$\theta$ is the angle of deflection.
}
\label{fig:geometry}
\end{center}
\end{figure}

The distribution of resulting directions emerging from the collision is
an interesting quantity to examine. The angle of deflection $\theta$
(the inclination angle) after a collision can be anisotropic, but the
azimuthal angle $\phi$ must always be isotropic. We use the convention
that $\pi/2 > \theta  \ge 0 $ is the direction of forward scattering.
This is shown in Fig. \ref{fig:geometry}

When two chains collide, they remain in contact for some period of time,
$t_c$, that depends on the precise details of the initial configurations
of the chains. Because the chains are flexible, we expect to have highly
inelastic collisions.  In the limit of a very long $t_c$, much longer than
the relaxation of a chain, the two chains will have fully thermalized.
We will now describe what we expect in this limit.

\subsection{Thermal Limit}

When the chains are in contact for long enough that they have fully equilibrated,
the energy in the velocity degrees of freedom will then be described by the equipartition theorem~\cite{Reif}
with the caveat that they strictly obey conservation of energy, momentum, and angular momentum. For 
a large number of monomers, momentum and angular momentum conservation make a negligible correction to the energy
in each degree of freedom. The initial energy is
\begin{equation}
E = \frac{1}{2m}  {P_I}^2 (\frac{1}{N_1} + \frac{1}{N_2})
\end{equation}
Assuming thermal equilibrium, when the two chains collide, we can define a temperature $T$, and statistical
mechanics gives the relationship between this and the energy $E$. 
The average total kinetic energy is $\langle K \rangle = (N_1+N_2) d_f k_B T/2$, where $d_f$ are the number of degrees of freedom
per monomer, so in the special case of an athermal system, $E = K$ so that 
\begin{equation}
\label{eq:athermalT}
k_B T = \frac{{P_I}^2}{d_f m}  \frac{(\frac{1}{N_1} + \frac{1}{N_2})}{N_1+N_2} = \frac{{P_I}^2}{d_f m(N_1 N_2)}  
\end{equation}

We would like to calculate the average center of mass energy of both molecules after the collision.
The position and momentum degrees of freedom are independent so we need only consider kinetic
energy corresponding to particles with momenta ${\bf p}_{j,1}$, and  ${\bf p}_{j,2}$ where the first subscript, $j$,
indexes the monomer, and the second subscript indexes the polymer.
\begin{equation}
K = \frac{1}{2m}(\sum_{j=1}^{N_1} p_{j,1}^2 + \sum_{j=1}^{N_2} p_{j,2}^2)
\end{equation}
The probability of finding the system with a given set of momenta is proportional to $\exp(-\beta K)$.
subject to the constraint the 
\begin{equation}
\sum_{j=1}^{N_1} {\bf p}_{j,1} + \sum_{j=1}^{N_2} {\bf p}_{j,2} = 0
\end{equation}
We would like to find the probability distribution of the total momemtum of one chain
\begin{equation}
{\bf P}_1 = \sum_{j=1}^{N_1} {\bf p}_{j,1}
\end{equation}
This problem is mathematically identical to the problem of a gaussian ring polymer with $N_1 + N_2$ monomers. In particular,
one can regard the particle momenta as displacement vectors between nearest neighbors.
Then ${\bf P}_1$ can be interpreted as the displacement of two monomers
separated by $N_1$ monomers. For the purposes of this calculation, this is the same as two linear chains in parallel, of lengths
$N_1$ and $N_2$. Therefore probability distribution of ${\bf P}_1$ is 
\begin{equation}
\label{eq:P(P1)}
P({\bf P}_1) \propto e^{-\frac{\beta {P_1}^2}{2} (\frac{1}{M_1} + \frac{1}{M_2})}
\end{equation}
Therefore 
\begin{equation}
\label{eq:P1sq=3kbT}
\langle P_1^2 \rangle = 3 k_B T M_r
\end{equation}
where $1/M_r = 1/M_1 + 1/M_2$ is the reduced mass.

Alternatively this result can be seen by considering a gas in thermal
equilibrium and considering the joint momentum distribution of two molecules of
mass $M_1$ and $M_2$, $P({\bf p}_1, {\bf p}_2) \propto \exp(-\beta K)$. 
We are interested in the distribution of momentum
in the center of mass reference frame. The kinetic energy $K$ is the sum of
the internal plus center of mass kinetic energy. Because the internal
momenta are equal and opposite, the internal term $K_I$
is 
\begin{equation}
K_I = \frac{{P_1}^2}{2}(\frac{1}{M_1} + \frac{1}{M_2})
\end{equation}
and is independent of the center of mass momentum. This implies
the distribution of $P_1$ is of the form of Eq. \ref{eq:P(P1)}.

In the athermal limit, we can use the value of $T$ given by Eq. \ref{eq:athermalT} and substitute that into Eq. \ref{eq:P1sq=3kbT}
obtaining
\begin{equation}
\langle {P_1}^2 \rangle = 3 \frac{{P_I}^2}{d_f (N_1+N_2)} .
\end{equation}
Therefore the variance of the final center of mass speed of chain 1, $v_{cm,1,F}$, is
related to its initial speed $\vcmi$ as
\begin{equation}
\label{eq:ave_vcm}
\langle {v_{cm,1,F}}^2 \rangle = 3 \frac{\vcmi^2}{d_f (N_1+N_2)}
\end{equation}


In a gas of molecules with a large number of internal degrees of freedom, the
distribution of total molecular energy is highly peaked. The standard deviation
of this energy is smaller than the mean by a factor of $1/\sqrt{N}$. In the
limit of large $N$ we can ignore ignore fluctuations in this energy. In thermal
equilibrium, we would like to know the distribution of center of mass speeds seen
that result from a collision. 

To derive this distribution, assume we have just two molecules
in a container with periodic boundary conditions with total center of mass
momentum of zero. Assume the size of the container is much larger than the size 
of a molecule and that the system is in thermal equilibrium. Such a system is expected
to be ergodic and hence be able to reach thermal equilibrium.
The distribution of the center of mass speed measured at one time is Maxwellian. This differs from
the distribution of speeds that are measured after a collision. In the latter case, we are not
measuring speeds at arbitrary times, but rather, after 
a collision. This difference in sampling alters the distribution. 
After a collision the center of mass of molecule $1$, will move in a straight line with speed $v_{cm,1,F}$  
until it suffers another collision.  The time spent in this state before 
undergoing another collision will be $\propto 1/v_{cm,1,F}$. So to compenstate for this, the distribution
of final speeds must be multiplied by an extra factor of $ v_{cm,1,F}$ so that
\begin{equation}
\label{eq:Pvf_therm}
P(v_{cm,1,F}) dv_{cm,1,F} \propto v_{cm,1,F}^3 e^{-\frac{v_{cm,1,F}^2}{2\sigma_x^2}} dv_{cm,1,F}.
\end{equation}
Here $\sigma_x^2$ is the variance of one of the components of the final velocity, and therefore
$\sigma_x^2 = \langle v_{cm,1,F}^2\rangle/3$, or using Eq. \ref{eq:ave_vcm},
\begin{equation}
\label{eq:sigma_x_v_cm}
\sigma_x^2 =  \frac{\vcmi^2}{d_f (N_1+N_2)}
\end{equation}

For a gas in thermal equilibrium, the distribution
of speeds of molecules striking a wall is well known from the study of effusion~\cite{ReifEffusion}
and this is the form that is expected in that case as well.

The distribution of directions for the final velocity in this limit should be isotropic. The
only subtlety being that the total angular momentum $L_t$ is conserved, so if $L_t$ is zero,
it might be thought that this would preclude the two chains rotating relative to each other
and that this would cause them separate in a way that preserves their initial relative
orientation. However it is still possible to rotate the two chains by 
circular motion of only parts of them. For example, if an end monomer rotates by one revolution,
this will cause the rest of the system to compensate by rotating in the opposite direction. This
will cause a rotation in the relative positions of the two chains, without changing the total
angular momentum.
Therefore in the limit of long collision times, the relative orientation of the two chains is not 
conserved and the final direction should be isotropic.

In terms of the variable
\begin{equation}
\label{eq:zcos}
z \equiv \cos \theta
\end{equation}
the distribution of $z$ should be uniform. For this reason we will use $z$ rather than $\theta$ to characterize
the results of the simulations discussed below.

Note that the above consideration could apply to any large molecules undergoing collisions, not
just polymers. The only thing required is that they remain in contact for sufficiently long
that thermal equilibrium is established.

\subsection{Simulation}

We model self avoiding chains as was done previously~\cite{DeutschVacPRL}. The distance
between links is maintained at a constant value of $l = 1$. The chain is modeled as freely hinged and
there is no chain stiffness.  A repulsive potential between all monomers was included. If the
distance between two monomers is $r$, their potential was taken to be 
$V(r) = 40(1-(r/l)^2)^5$. For $r > l $, $V(r) = 0$. A diverging hard core was not used for 
the sake of efficiency. The potential at the center is very high making chain crossing exceedingly
unlikely and no chain crossing was observed. 

Rigid links were chosen to avoid equilibration problems that can occur due to the quasi-one dimensional
nature of this system~\cite{FPU,BermanIzrailev}. The simulation method solves
Newton's laws for this system so that all conservation laws are well satisfied.
It also implements the rigid link constraints in an efficient way, using 
$O(N)$ operations for every integration step. The method is described in detail
in Ref. \cite{DeutschCerf}. 

Many collisions were studied with different initial configurations of the chains and their statistics
were analyzed. We took the initial center speed of chains to be $1$, and $m=1$.
The time step for the simulations was $0.02$. 

The chains were initially equilibrated for many relaxation times. For example for $N=128$
the equilibration time was $40,000$ steps. Then their velocities were made
negligibly small by applying a large damping for many damping times so that their final velocities 
were less than $10^{-8}$. These chains were given their initial center of mass speeds $\vcmi = 1$
in the direction between the two center of masses. This way there was no initial internal energy
of the chains, corresponding to the limit of high relative collision velocity. The chains were initially
separated by at least half of their total chain length. This ensures that they are well separated before
a collision takes place. To obtain adequate statistics, many collisions, about  $10^4$, were used 
with different random initial conditions.

The length of time they remained in contact was determined by monitoring
how the center of mass velocity $\vcm$ changed. Before a collision, as the chains
approach each other, $\vcm$ remains constant, but then
as soon as the chains collide, there will be a change in this velocity. The point
where $\vcm$ first changes signals the start of a collision. In order for a collision to be
considered over, $\vcm$ had to remain constant for
all subsequent times. Time evolution was not stopped until the chains
had completely separated from each. This was implemented by requiring that
the evolution continue if there were two monomers from different chains
that were closer than half their original separation at the start of the
simulation. This eliminated the cases where the two chains appeared to
have separated but later collided again.
The final speeds were also recorded, as was the angle of deflection $\theta$ of the chains
after the collision (again, with the convention that $ \pi/2 > \theta \ge 0$ is forward scattering).

\subsection{Results}

\begin{figure}[htp]
\begin{center}
\includegraphics[width=\hsize]{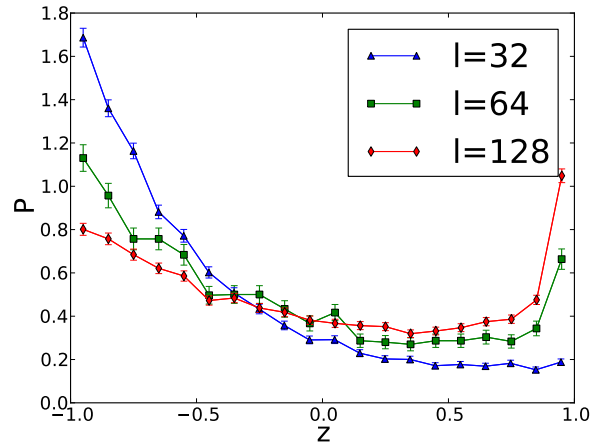}
\caption{ (Color Online) 
The distribution of scattering angles $\theta$. The horizontal axis
is $z=\cos \theta$. The highest curve on the left is for $N = 32$,
the middle $N=64$, and the lowest $N=128$. 
}
\label{fig:Pangle}
\end{center}
\end{figure}

Fig. \ref{fig:Pangle} shows the distribution of scattering angles as a function
of $z$, as defined in Eq. \ref{eq:zcos}. For  $N = 32$, there is more scattering 
backwards, that is, in the direction of negative $z$. For larger $N$, this changes
and for $N=128$, the effect of backwards scattering is much weaker. Instead, there
is a strong peak for $z=1$. This means that a substantial fraction of collisions deflect the
polymers by a very small amount. We will discuss this further below.

\begin{figure}[htp]
\begin{center}
\includegraphics[width=\hsize]{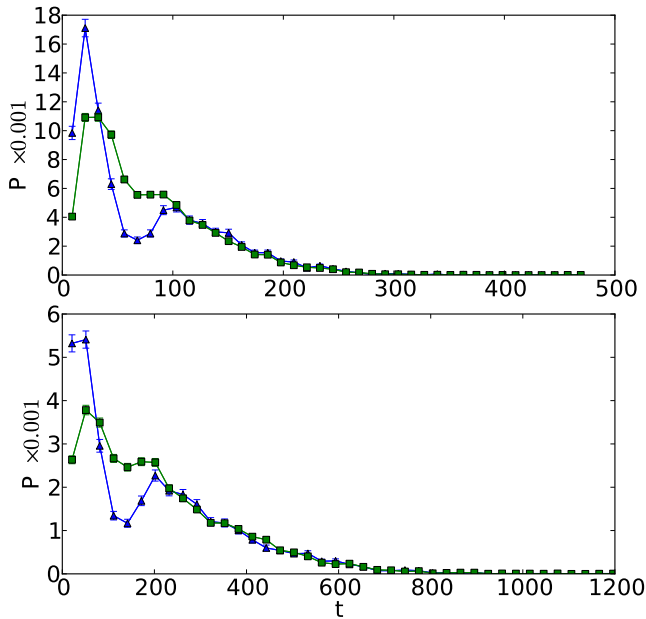}
\caption{ (Color Online) 
The distribution of collision times $t$ for a chain with $N=128$ (lower graph)
and $N=64$ (upper graph),
and the same distribution excluding those where the scattering angle is
greater than $\pi/2$. The latter distribution has a sizable dip at $t \approx 150$ for $N=128$.
}
\label{fig:t_2hists}
\end{center}
\end{figure}

Assuming an athermal model, the temperature given by Eq.  \ref{eq:athermalT} is $k_B T = 1/2$.
The distribution of collision times is shown in Fig. \ref{fig:t_2hists} for $N = 128$ (lower
graph) and $N = 64$ (upper graph.) The average time computed from this distribution, for $N=128$, is $210$.
This is comparable but larger than the equilibrium relaxation time $t_{rel}$ seen for
similar simulations of equilibrium systems. 
The latter was determined from measuring the time dependent autocorrelation function as
has been described previously, see Ref. \cite{DeutschVacPRL}, Fig. 3.
This suggests that during the majority of these collisions, the system should be
close to thermal equilibrium.
There are two sets of data shown on each plot in Fig. \ref{fig:t_2hists}. One where all scattering angles are included (triangles), the
second only includes forward scattering, $z > 0$ (squares). With only forward scattering,
there is a prominent dip in the distribution at $t \approx 150$ ($N = 128$.)  For longer times,
the two curves converge quite closely. The reason for this unusual small $t$ behavior
will also be discussed below.

\begin{figure*}[htp]
\begin{center}
(a)
\includegraphics[width=.4 \hsize]{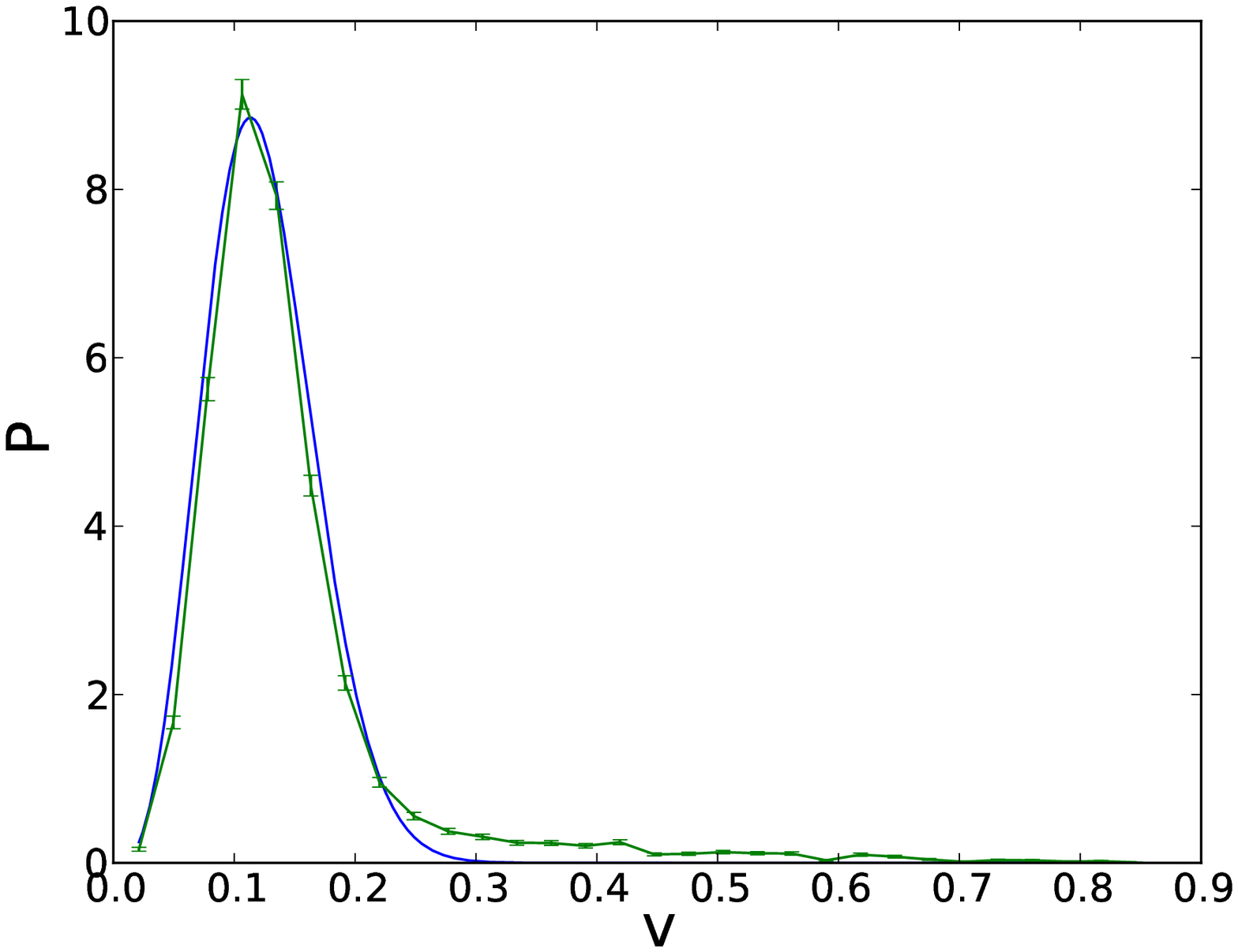}
(b)
\includegraphics[width=.4 \hsize]{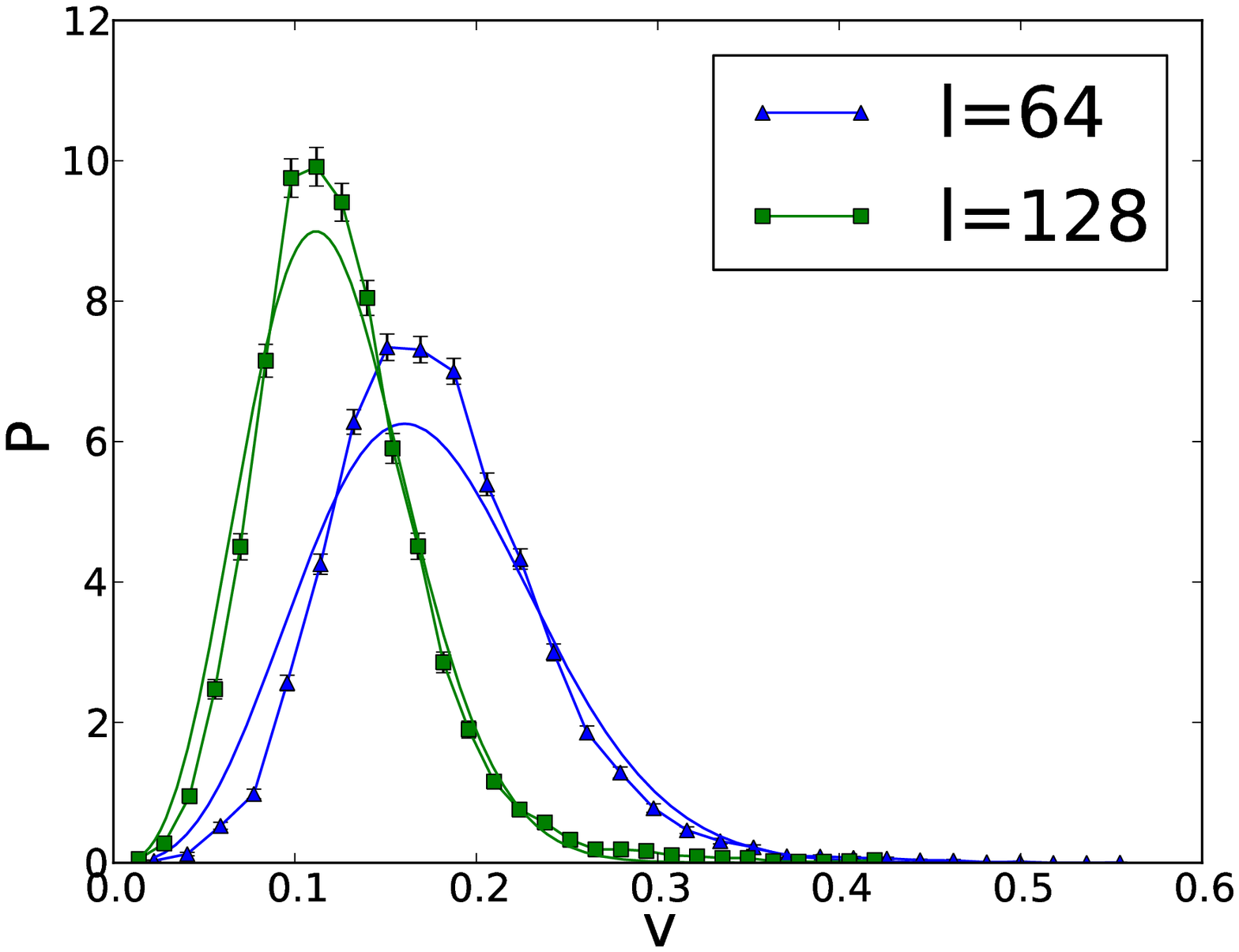}
\caption{ (Color Online) 
(a) The distribution of final collision speeds $v$ including all scattering angles
and $N = 128$.
(b) The same distribution ignoring those where $\cos \theta < 0.9$. For comparison the
distribution for $L=64$ is also shown.
}
\label{fig:Pv}
\end{center}
\end{figure*}

The distribution of final velocities is shown in Fig. \ref{fig:Pv}(a). The smooth curve
without error bars, is a fit to the form expected in the thermal limit, Eq.  \ref{eq:Pvf_therm}.
Note that there is a much longer tail in the simulation data. If instead one excludes strong
forward scattering, $z > 0.9$, as shown in Fig. \ref{fig:Pv}(b), the fit to the thermal
distribution is much improved. The fit gives $\sigma_x = 0.064$, where as the predicted
value from Eq. \ref{eq:sigma_x_v_cm}, the thermal limit, is $0.044$.
The distribution in  Fig. \ref{fig:Pv}(b) is still slightly too wide and this suggests that
low final velocities are suppressed compared to the thermal prediction.

\begin{figure}[htp]
\begin{center}
\includegraphics[width=\hsize]{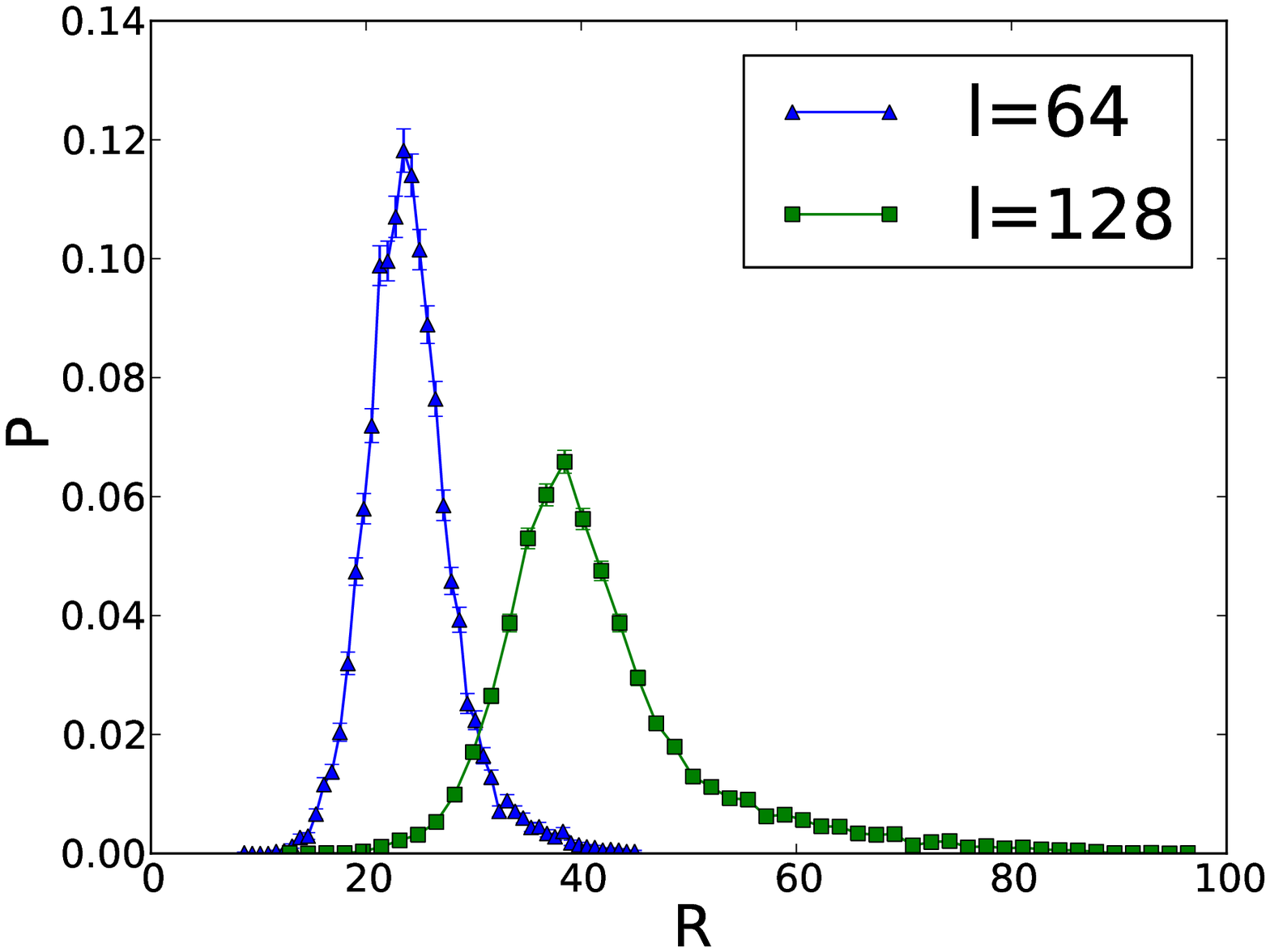}
\caption{ (Color Online) 
The distribution of the maximum end to end distances $R$, occurring during scattering for
chains with $N = 64$ and $128$.
}
\label{fig:Pendtoend}
\end{center}
\end{figure}

The distribution of the maximum end to end distances $R$, occurring during scattering is shown in
Fig. \ref{fig:Pendtoend} for $N = 64$ and $N = 128$. There is typically substantial stretching that occurs
during a collision. For comparison, statistics for the end to end distance $R_e$ for $N = 128$ were calculated
in thermal equilibrium for a total angular momentum of zero. The same chain has an rms end to end separation of
$\sqrt{\langle R_e^2\rangle} = 17.5$,
However the maximum in Fig. \ref{fig:Pendtoend} is at $R=38$ and has a significant tail stretching to
past $R=90$. After the chains have separated, the stretching will disappear and tend towards
their equilibrium values.

\begin{figure*}[htp]
\begin{center}
(a)
\includegraphics[width=.45 \hsize]{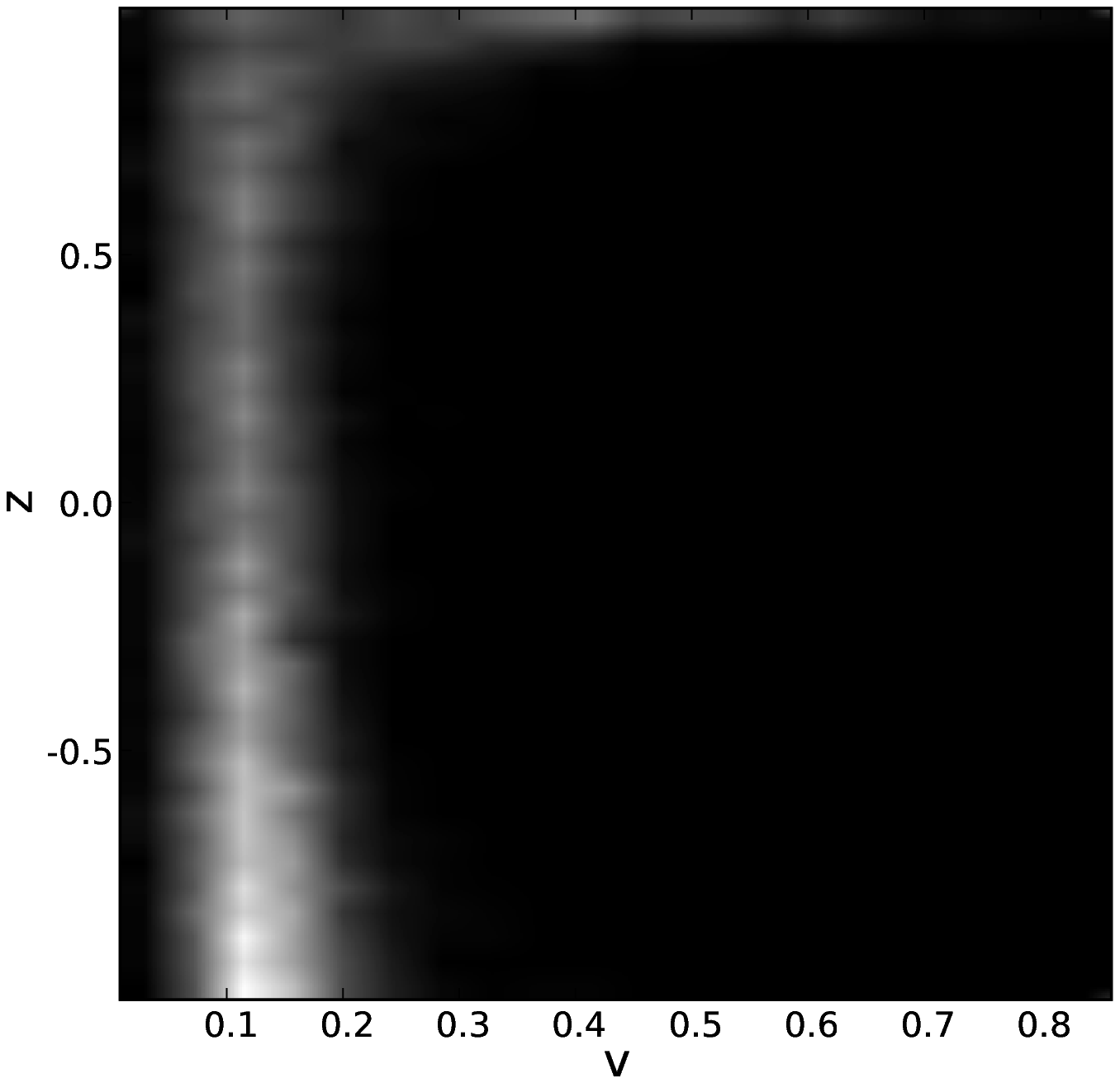}
(b)
\includegraphics[width=.45 \hsize]{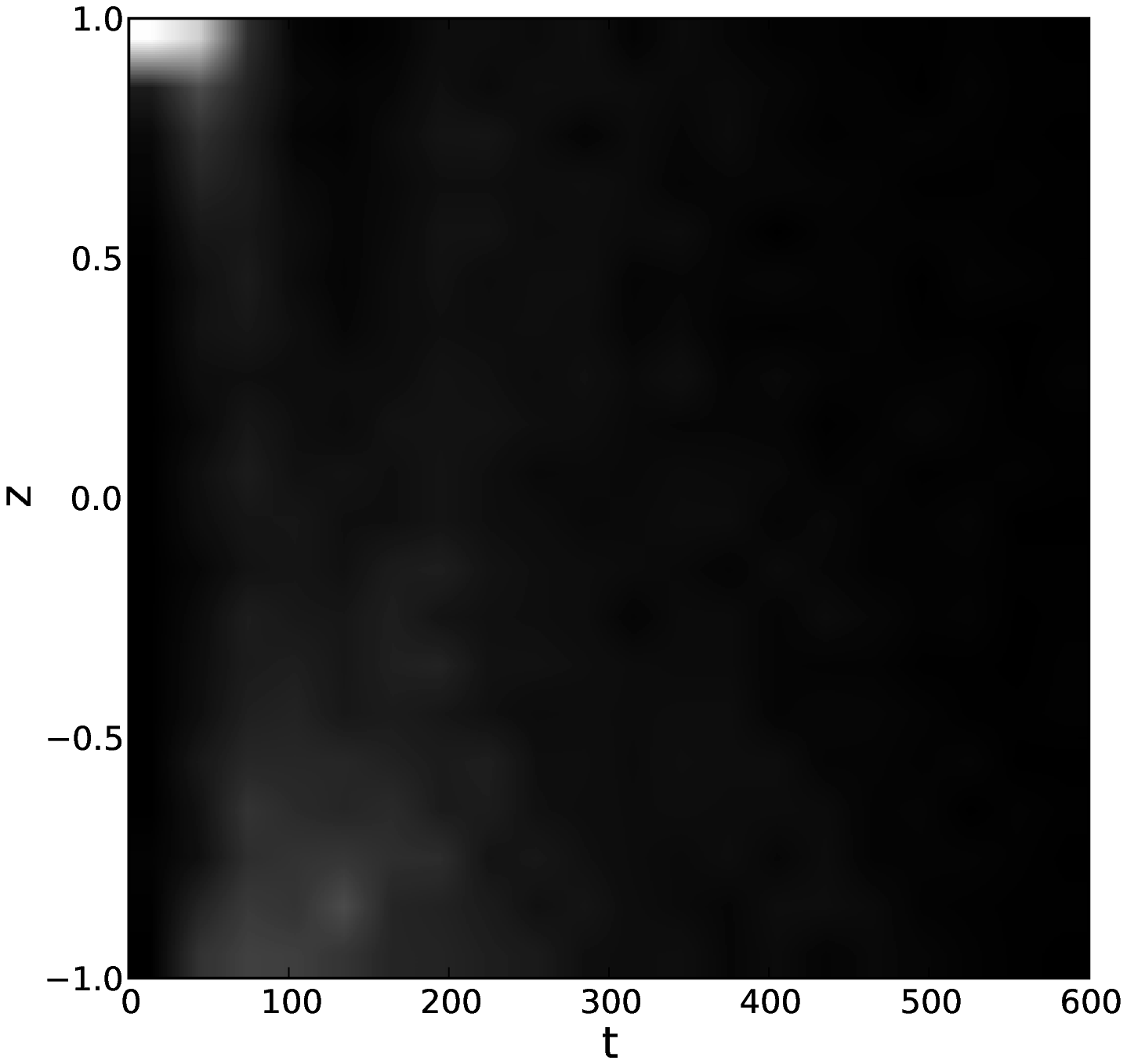}
\caption{ 
(a) The probability density as a function of final speed $v$, and
the amount of deflection $z = \cos \theta$.
(b)
The probability density as a function of collision time $t$, and
the amount of deflection $z = \cos \theta$.
}
\label{fig:2d_cos_v}
\end{center}
\end{figure*}

To summarize what we have learned so far, collisions between self avoiding chains in a vacuum appear
be quite close to what one would expect in the thermal limit. For the longest chain studied, $N=128$,
the scattering is close to isotropic with the exception of a spike for strong forward scattering. The
time that the chains remain in contact is typically longer than a relaxation time. The distribution
of final speeds appears quite comparable to what one would expect from a thermal distribution except for
collisions coming from strong forward scattering. Finally, at some point while the chains are in a collision,
they are typically quite stretched.

From this it appears that these chain collisions are highly inelastic and they have properties close to that
of collisions in the thermal limit. However there are some interesting deviations from this limit as was
noted above. There appears to be a fraction of collision where the angle of scattering is very
small. If these chains are excluded, then the collisions have statistics much closer to that of the thermal
limit. A simple explanation for this would be that this sub-population of chains are not actually colliding. However
their final velocities are substantially different from their initial ones, so this effect
is clearly more subtle. By observing such strong forward scattering collisions it appears that their
behavior can be understood as follows.

The dynamics of chains during collisions can be understood in terms of
two factors, entanglement effects and transfer of momentum. During strong
forward scattering collisions, it appears from visualizations, that the
chains do strongly collide but only weakly entangle.  Different parts of
the chains come into contact and when they do, these parts collide very
inelastically. That is, they do not bounce off one another but remain in
contact for some length of time. Therefore momentum transfer between the
chains is still inelastic. If two inelastic masses collide, even with
unequal mass, their final velocity will still be along the same line
as the initial velocities. Therefore all such inelastic collisions do
not change the final direction. Because the chains have not entangled,
they move past each other, but have not remained in contact long enough
to thermalize. Their final directions are almost unchanged but their
speeds have been substantially reduced due to momentum transfer.

The above observations can be seen by examining two dimensional distributions, for $N = 128$ in  Fig. \ref{fig:2d_cos_v}.
In Fig. \ref{fig:2d_cos_v}(a), the distribution of collisions is binned in terms of final collision
speed $v$, and $z$. One can see that for strong forward scattering, $z \approx 1$, there are a large
number of collisions with final speeds much larger than the thermal velocity. In 
In Fig. \ref{fig:2d_cos_v}(b), the distribution of collisions is binned as a function of collision time
$t$ and $z$. Again for strong forward scattering, there is a high peak for atypically short times, 
indicating that the collisions are weaker than for other scattering angles. Therefore these
strong forward scattering collisions are short and do not slow down the chains all the way to their
thermal values. 

The plots in Figs. \ref{fig:t_2hists} and \ref{fig:Pv} are obtainable
from these two dimensional distributions. The enhancement in short
time collision frequency seen in Fig. \ref{fig:t_2hists} for $z > 0$ is
therefore caused by the subset of strong forward scattering collisions,
discussed above, that only collide for short times. All other collisions
that more completely thermalize, usually remain in contact for longer times. The
fact that there is a dip in the distribution of Fig. \ref{fig:t_2hists}
when considering collisions with $z > 0$ suggests that the quasi-thermal
chains having $z > 0$ have a peak in their collision time distribution at 
$t \approx 200$.

\section{Discussion}

It is of interest to examine the scaling of collisions with $N$. The radius of gyration of
a swollen chain $R \sim N^\nu$ with $\nu \approx 3/5$ in three dimensions. According to our assumption
that the initial kinetic energy of the chains are zero, the probability that a monomer from chain A
will collide with chain  B is obtained by looking at the projection of the density of chain B
into a two dimensional plane. The average density of this projection is proportional to the
probability of a collision. The projected density is proportional to  $N/R^2 \sim N^{1-2\nu} \approx N^{-1/5}$.
Therefore as $N\rightarrow \infty$ the probability that a given monomer in chain A will collide with
chain B goes to zero. On the other hand, the total number of contact points in chain A is $N^{4/5}$. 
Therefore for large $N$ the molecules will suffer a collision but the above argument suggests that
the number of monomers actually making contact will be a small fraction of the total. 

If the amount of time that these points remain in contact is not long enough, the chains will 
pass each other at some reduced center of mass velocities. If the collisions of contact points are
highly inelastic, only the speed of the chains, but not their direction, will change. 
Entanglements will change the above arguments considerably. Only one entanglement between two
chains can cause their relative velocity to go to zero. For long enough chains, one would expect
that entanglements would then dominate the collisions. However their effects usually only 
become dominant for chains larger than $N \approx 500$, as has been seen with the study of
topological effects in flexible polymer rings~\cite{DeutschRings}

To understand
the details of the dynamics of these collisions analytically is particularly difficult in
light of the topological nature of the interactions and is beyond the scope of the present work.
I would conjecture that the effects of entanglements will strongly suppress these forward 
scattering collisions for long enough chains.

If the chains are of different sizes, then their behavior will depend 
on their relative sizes. From the above discussion, the probability that a
single monomer will collide with a polymer of $N_1$ links is $\propto N_1^{-1/5}$.
Therefore if a second polymer is of length $N_2 < N_1$, the probability of one
collision is approximately $\propto N_2/N_1^{1/5}$. In other words, the second molecule is unlikely
to collide if $N_2 \ll N_1^{1/5}$. For chains of comparable sizes, the above
results suggest that collisions will still be strongly inelastic and will also
show behavior close to the thermal limit. The departure from this
due to forward scattering will increase as the mass difference increases.

\section{Conclusions}

In this work, collisions between polymers in a vacuum have been examined.
This is quite an unusual situation because collisions typically occur
between much smaller molecules, and consequently new features of this
situation are expected.  The focus here is on chains where the initial
relative center of mass velocity is so large that the initial internal
thermal energy of the molecule can be ignored. This is an interesting
limit to consider because for athermal chains, the magnitude of the
initial velocity factors out.

Typically, in a collision of such long molecules, they remain in contact
for long enough to be close to equilibrium so that they can be described
by thermal averages. The final velocities and scattering angles, are close
to such averages. However the main discrepancy with this description is
due to a class of collisions that strongly forward scatter and remain
in contact for a relatively short amount of time. This is evidence of
strongly inelastic collisions of subsections of the chains. It is expected
that for long enough chains, $N > 500$ entanglement effects will diminish
the frequency of these kinds of collisions.

In future work, it would be interesting to consider the case of chains with
strong attractive forces as this may also be relevant to experiments and
has interesting physical consequences. For example, chains at temperatures
that are initially below the coil-globule transition will tend to stick together for
low enough relative velocities. However beyond a threshold velocity, they would
be expected to heat enough upon impact to become swollen and then separate from
each other.

\end{document}